\documentstyle[12pt,epsf]{article}
\newcommand{\nc}{\newcommand}
\nc{\renc}{\renewcommand}

\renc{\baselinestretch}{1.2}
\nc{\com}[1]{\\{\bf \# {#1}}\\}
\newlength{\overeqskip}
\newlength{\undereqskip}
\nc{\be}[1]{\begin{equation} \mbox{$\label{#1}$}}
\nc{\bea}[1]{\begin{eqnarray} \mbox{$\label{#1}$}}
\nc{\Section}[2]{\section{\sc #2}\label{#1}}
\nc{\Bibitem}[1]{\bibitem{#1}}
\nc{\Label}[1]{\label{#1}}

\nc{\eea}{\vspace{\undereqskip}\end{eqnarray}}
\nc{\ee}{\vspace{\undereqskip}\end{equation}}
\nc{\bdm}{\begin{displaymath}}
\nc{\edm}{\end{displaymath}}
\nc{\dpsty}{\displaystyle}
\nc{\bc}{\begin{center}}
\nc{\ec}{\end{center}}
\nc{\ba}{\begin{array}}
\nc{\ea}{\end{array}}
\nc{\bab}{\begin{abstract}}
\nc{\eab}{\end{abstract}}
\nc{\btab}{\begin{tabular}}
\nc{\etab}{\end{tabular}}
\nc{\bit}{\begin{itemize}}
\nc{\eit}{\end{itemize}}
\nc{\ben}{\begin{enumerate}}
\nc{\een}{\end{enumerate}}
\nc{\bfig}{\begin{figure}}
\nc{\efig}{\end{figure}}
\nc{\eqs}[2]{\mbox{Eqs.(\ref{#1},\,\ref{#2})}}
\nc{\eq}[1]{\mbox{Eq.(\ref{#1})}}

\nc{\figs}[2]{\mbox{Figs.(\ref{#1},\,\ref{#2})}}
\nc{\fig}[1]{\mbox{Fig.(\ref{#1})}}

\nc{\figcap}[1]{\refstepcounter{figure}
	{\bf Figure \thefigure}: {\small\sl #1}}
\nc{\tabcap}[1]{\refstepcounter{table}
	{\bf Table \thetable}: {\small\sl #1}}

\nc{\tag}[1]{\label{#1} \marginpar{{\footnotesize #1}}}
\nc{\mtag}[1]{\label{#1} \mbox{\marginpar{{\footnotesize #1}}}}

\nc{\etal}{\mbox{\it et al. }}
\nc{\ie}{{\it i.e.}}
\nc{\eg}{{\it e.g.}}

\nc{\arreq}{&\!\!=\!\!&}
\nc{\arrmi}{&\!\!-\!\!&}
\nc{\arrpl}{&\!\!+\!\!&}
\nc{\arrap}{&\!\!\!\approx\!\!\!&}
\nc{\non}{\nonumber}
\nc{\align}{\!\!\!\!\!\!\!\!&&}

\def\lsim{\; \raise0.3ex\hbox{$<$\kern-0.75em
      \raise-1.1ex\hbox{$\sim$}}\; }
\def\gsim{\; \raise0.3ex\hbox{$>$\kern-0.75em
      \raise-1.1ex\hbox{$\sim$}}\; }
\nc{\DOT}{\hspace{-0.08in}{\bf .}\hspace{0.1in}}
\nc{\Laada}{\hbox {$\sqcap$ \kern -1em $\sqcup$}}
\nc\loota{{\scriptstyle\sqcap\kern-0.55em\hbox{$\scriptstyle\sqcup$}}}
\nc\Loota{{\sqcap\kern-0.65em\hbox{$\sqcup$}}}
\nc\laada{\Loota}
\nc{\qed}{\hskip 3em \hbox{\BOX} \vskip 2ex}

\nc{\real}{{\rm I \! R}}
\nc{\Z}{{\sf Z \!\!\! Z}}
\nc{\complex}{{\rm C\!\!\! {\sf I}\,\,}}
\def\bigid{\leavevmode\hbox{\small1\kern-3.8pt\normalsize1}}
\def\id{\leavevmode\hbox{\small1\kern-3.3pt\normalsize1}}
\nc{\slask}{\!\!\!/}
\nc{\bis}{{\prime\prime}}
\nc{\pa}{\partial}
\nc{\na}{\nabla}
\nc{\ra}{\rangle}
\nc{\la}{\langle}
\nc{\goto}{\rightarrow}
\nc{\swap}{\leftrightarrow}

\nc{\EE}[1]{ \mbox{$\cdot10^{#1}$} }
\nc{\abs}[1]{\left|#1\right|}
\nc{\at}[2]{\left.#1\right|_{#2}}
\nc{\norm}[1]{\|#1\|}
\nc{\abscut}[2]{\Abs{#1}_{\scriptscriptstyle#2}}
\nc{\vek}[1]{{\rm\bf #1}}
\nc{\integral}[2]{\int\limits_{#1}^{#2}}
\nc{\inv}[1]{\frac{1}{#1}}
\nc{\dd}[2]{{{\partial #1}\over{\partial #2}}}
\nc{\ddd}[2]{{{{\partial}^2 #1}\over{\partial {#2}^2}}}
\nc{\dddd}[3]{{{{\partial}^2 #1}\over
	{\partial #2 \partial #3}}}
\nc{\dder}[2]{{{d #1}\over{d #2}}}
\nc{\ddder}[2]{{{d^2 #1}\over{d {#2}^2}}}
\nc{\dddder}[3]{{d^2 #1}\over
	{d #2 d #3}}
\nc{\dx}[1]{d\,^{#1}x}
\nc{\dy}[1]{d\,^{#1}y}
\nc{\dz}[1]{d\,^{#1}z}
\nc{\dl}[1]{\frac{d\,^{#1}l}{(2\pi)^{#1}}}
\nc{\dk}[1]{\frac{d\,^{#1}k}{(2\pi)^{#1}}}
\nc{\dq}[1]{\frac{d\,^{#1}q}{(2\pi)^{#1}}}

\nc{\cc}{\mbox{$c.c.$ }}
\nc{\hc}{\mbox{$h.c.$ }}
\nc{\cf}{cf.\ }
\nc{\erfc}{{\rm erfc}}
\nc{\Tr}{{\rm Tr\,}}
\nc{\tr}{{\rm tr\,}}
\nc{\pol}{{\rm pol}}
\nc{\sign}{{\rm sign}}
\nc{\bfT}{{\bf T }}

\nc{\cA}{{\cal A}}
\nc{\cB}{{\cal B}}
\nc{\cD}{{\cal D}}
\nc{\cE}{{\cal E}}
\nc{\cF}{{\cal F}}
\nc{\cG}{{\cal G}}
\nc{\cH}{{\cal H}}
\nc{\cL}{{\cal L}}
\nc{\cO}{{\cal O}}
\nc{\cT}{{\cal T}}
\nc{\rvac}[1]{|{\cal O}#1\rangle}
\nc{\lvac}[1]{\langle{\cal O}#1|}
\nc{\rvacb}[1]{|{\cal O}_\beta #1\rangle}
\nc{\lvacb}[1]{\langle{\cal O}_\beta #1 |}
\nc{\bb}{\bar{\beta}}
\nc{\ctH}{\tilde{\cal H}}
\nc{\chH}{\hat{\cal H}}
\nc{\1}{\aa}
\nc{\2}{\"{a}}
\nc{\3}{\"{o}}
\nc{\4}{\AA}
\nc{\5}{\"{A}}
\nc{\6}{\"{O}}
\nc{\al}{\alpha}
\nc{\Del}{\Delta}
\nc{\e}{\epsilon}
\nc{\eps}{\epsilon}
\nc{\lam}{\lambda}
\nc{\om}{\omega}
\nc{\Om}{\Omega}
\nc{\ve}{\varepsilon}
\nc{\mn}{{\mu\nu}}
\nc{\k}{\kappa}
\nc{\vp}{\varphi}

\nc{\advp}[3]{{\it  Adv.\ in\ Phys.\ }{{\bf #1} {(#2)} {#3}}}
\nc{\annp}[3]{{\it  Ann.\ Phys.\ (N.Y.)\ }{{\bf #1} {(#2)} {#3}}}
\nc{\apl}[3]{{\it  Appl. Phys. Lett. }{{\bf #1} {(#2)} {#3}}}
\nc{\apj}[3]{{\it  Ap.\ J.\ }{{\bf #1} {(#2)} {#3}}}
\nc{\apjl}[3]{{\it  Ap.\ J.\ Lett.\ }{{\bf #1} {(#2)} {#3}}}
\nc{\app}[3]{{\it Astropart.\ Phys.\ }{{\bf #1} {(#2)} {#3}}}
\nc{\cmp}[3]{{\it  Comm.\ Math.\ Phys.\ }{{ \bf #1} {(#2)} {#3}}}
\nc{\cqg}[3]{{\it  Class.\ Quant.\ Grav.\ }{{\bf #1} {(#2)} {#3}}}
\nc{\epl}[3]{{\it  Europhys.\ Lett.\ }{{\bf #1} {(#2)} {#3}}}
\nc{\ijmp}[3]{{\it Int.\ J.\ Mod.\ Phys.\ }{{\bf #1} {(#2)} {#3}}}
\nc{\ijtp}[3]{{\it Int.\ J.\ Theor.\ Phys.\ }{{\bf #1} {(#2)} {#3}}}
\nc{\jmp}[3]{{\it  J.\ Math.\ Phys.\ }{{ \bf #1} {(#2)} {#3}}}
\nc{\jpa}[3]{{\it  J.\ Phys.\ A\ }{{\bf #1} {(#2)} {#3}}}
\nc{\jpc}[3]{{\it  J.\ Phys.\ C\ }{{\bf #1} {(#2)} {#3}}}
\nc{\jap}[3]{{\it J.\ Appl.\ Phys.\ }{{\bf #1} {(#2)} {#3}}}
\nc{\jpsj}[3]{{\it J.\ Phys.\ Soc.\ Japan\ }{{\bf #1} {(#2)} {#3}}}
\nc{\lmp}[3]{{\it Lett.\ Math.\ Phys.\ }{{\bf #1} {(#2)} {#3}}}
\nc{\mpl}[3]{{\it  Mod.\ Phys.\ Lett.\ }{{\bf #1} {(#2)} {#3}}}
\nc{\ncim}[3]{{\it  Nuov.\ Cim.\ }{{\bf #1} {(#2)} {#3}}}
\nc{\np}[3]{{\it  Nucl.\ Phys.\ }{{\bf #1} {(#2)} {#3}}}
\nc{\pr}[3]{{\it Phys.\ Rev.\ }{{\bf #1} {(#2)} {#3}}}
\nc{\pra}[3]{{\it  Phys.\ Rev.\ A\ }{{\bf #1} {(#2)} {#3}}}
\nc{\prb}[3]{{\it  Phys.\ Rev.\ B\ }{{{\bf #1} {(#2)} {#3}}}}
\nc{\prc}[3]{{\it  Phys.\ Rev.\ C\ }{{\bf #1} {(#2)} {#3}}}
\nc{\prd}[3]{{\it  Phys.\ Rev.\ D\ }{{\bf #1} {(#2)} {#3}}}
\nc{\prl}[3]{{\it Phys.\ Rev.\ Lett.\ }{{\bf #1} {(#2)} {#3}}}
\nc{\pl}[3]{{\it  Phys.\ Lett.\ }{{\bf #1} {(#2)} {#3}}}
\nc{\prep}[3]{{\it Phys\. Rep.\ }{{\bf #1} {(#2)} {#3}}}
\nc{\prsl}[3]{{\it Proc.\ R.\ Soc.\ London\ }{{\bf #1} {(#2)} {#3}}}
\nc{\ptp}[3]{{\it  Prog.\ Theor.\ Phys.\ }{{\bf #1} {(#2)} {#3}}}
\nc{\ptps}[3]{{\it  Prog\ Theor.\ Phys.\ suppl.\ }{{\bf #1} {(#2)} {#3}}}
\nc{\physa}[3]{{\it  Physica\ A\ }{{\bf #1} {(#2)} {#3}}}
\nc{\physb}[3]{{\it  Physica\ B\ }{{\bf #1} {(#2)} {#3}}}
\nc{\phys}[3]{{\it Physica\ }{{\bf #1} {(#2)} {#3}}}
\nc{\rmp}[3]{{\it  Rev.\ Mod.\ Phys.\ }{{\bf #1} {(#2)} {#3}}}
\nc{\rpp}[3]{{\it Rep.\ Prog.\ Phys.\ }{{\bf #1} {(#2)} {#3}}}
\nc{\sjnp}[3]{{\it Sov.\ J.\ Nucl.\ Phys.\ }{{\bf #1} {(#2)} {#3}}}
\nc{\spjetp}[3]{{\it Sov.\ Phys.\ JETP\ }{{\bf #1} {(#2)} {#3}}}
\nc{\yf}[3]{{\it Yad.\ Fiz.\ }{{\bf #1} {(#2)} {#3}}}
\nc{\zetp}[3]{{\it Zh.\ Eksp.\ Teor.\ Fiz.\ }{{\bf #1} {(#2)} {#3}}}
\nc{\zp}[3]{{\it Z.\ Phys.\ }{{\bf #1} {(#2)} {#3}}}
\nc{\ibid}[3]{{\sl ibid.\ }{{\bf #1} {#2} {#3}}}
\nc{\rf}[1]{(\ref{#1})}
\nc{\nn}{\nonumber \\*}
\nc{\Lbmeff}{\cL^{\beta,\mu}_{\rm eff}}
\nc{\Fmn}{F_{\mu\nu}}
\nc{\bt}{\beta_\tau}
\nc{\bk}{\beta_\kappa}
\nc{\g}{g(T)}
\nc{\as}{\alpha_s}
\nc{\vdw}{Vilkovisky--DeWitt}
\nc{\trge}{temperature renormalization group equation}
\nc{\rge}{renormalization group equation}
\nc{\htl}{hard thermal loop}
\nc{\gtk}{g\tau/\kappa}
\nc{\gttk}{g^2\tau/\kappa}
\nc{\tk}{\tau/\kappa}
\nc{\Tk}{(T,\kappa)}
\nc{\tauk}{(\tau,\kappa)}
\begin{document}
%
\large
\thispagestyle{empty}
\begin{flushright}
	August, 1994\\
	NORDITA-94/36 P \\
        hep-ph/yymmnn
 \end{flushright}
\bc
{\Huge\bf The Thermal $\beta$-Function \\[6mm]
            in  Yang-Mills Theory}
\\[1cm]
\ec
\vspace*{1cm}
\bc
{\large{\bf Per Elmfors$^{\dagger,}$\footnote{
E-mail: elmfors@nordita.dk}}
and {\bf Randy Kobes$^{*,+,}$}\footnote{
E-mail: randy@theory.uwinnipeg.ca}}\\[5mm]
{\normalsize$^\dagger$NORDITA,
Blegdamsvej 17,
DK-2100 Copenhagen \O,
Denmark\\
$^*$ENSLAPP,
Chemin de Bellevue BP 110,
F-749 41 Annecy-le-Vieux Cedex,
France\\
$^+$Physics Dept., University of Winnipeg,
Winnipeg, MB\ R3B 2E9, Canada\\}
\ec
\
\vspace*{2cm}
\bc
{\bf Abstract} \\
\ec
{\normalsize
\begin{quotation}
\noindent
 Previous calculations of the thermal
$\beta$--function in a hot
Yang--Mills gas
at the one--loop level have exposed problems
with the gauge dependence and with the sign, which is opposite to
what one would expect for asymptotic freedom.
We show that inclusion
of higher--loop effects through a static Braaten--Pisarski resummation
is necessary to consistently obtain the leading term, but
alters the results only quantitatively. The sign, in particular,
remains the same. We also explore, by a crude parameterization, the
 effects a (non--perturbative) magnetic mass may
have on these results.
\end{quotation}}
\vfill
\newpage
%
\normalsize
\setcounter{page}{1}
\Section{intro}{Introduction}
%
The behaviour of the effective coupling constant $\as=g^2/4\pi$ in
QCD at high temperature or density has been discussed for a long
time, starting with the renormalization group equation (RGE)
arguments of Collins and Perry \cite{CollinsP75} that $\as$ decreases
logarithmically at high density due to asymptotic freedom.
The idea of QCD as a gas of weakly interacting quarks and gluons
at high $T$ originates from this observation.
It has later been questioned if it is correct to use the same decreasing
$\as$ as the renormalized coupling constant when computing
 general $n$--point
functions with non--zero external momenta, as the simple
scaling assumptions used in  \cite{CollinsP75}
do not hold when the external momenta introduce extra
dimensionful parameters.
The zero temperature RGE can only be expected to be useful
when the typical
momenta involved scale with the temperature \cite{Yamada88,EnqvistK92}. Also,
the argument in  \cite{CollinsP75} assumes that there
are no infra--red
problems, which are now known to exist \cite{Linde80}.
Therefore, several
groups have explicitly calculated the $T$--dependence
of the three--point function in QCD at high $T$ and used a \rge,
with the temperature and the external momentum $\kappa$ as
scale parameters \cite{MatsumotoNU84}, in
order to derive the running of $\as$ with $\Tk$
[6 -- 15].
Even if $\as$ was found to decrease logarithmically
at high $T$ it would not be
enough to justify an ideal gas approximation of QCD since
the typical expansion
parameters $\as T/\kappa$ and $\sqrt{\as}T/\kappa$ still
grow at high $T$.

Various problems and ambiguities arose when calculating the thermal
$\beta$--function. It was recognized soon that the
dependence of $\as\Tk$
on $T$ depends strongly on which vertex is chosen to
renormalize $\as$,
 the other vertices being determined by Ward identities
\cite{NakkagawaN87,FujimotoY88,NakkagawaNY88,BaierPS90,NakkagawaNY90}.
This prescription dependence exists also at T=0  when
the momentum--space
subtraction is used \cite{CelmasterG79}.
There was also some ambiguity in the results which
depended on whether the imaginary time formalism (ITF)
or a real time formalism was used
\cite{BaierPS90,NakkagawaNY90}, but this is now better
understood \cite{BaierPS91}.
Furthermore, for a given vertex the $\beta$-function depends on
the momentum prescription and differs, for example,
when the collinear and the symmetric
points are used, both at zero external energy.
Another problem arose in that the result is also gauge fixing
dependent \cite{NakkagawaNY88}, which puts into serious question
the usefulness of such an approach.
It is, in fact, not at all surprising that the $\beta$--function
 shows a gauge dependence when computed using the
standard effective action
\cite{NakkagawaNY88} since it is not
gauge invariant off--shell. Landsman therefore proposed \cite{Landsman89}
to use the \vdw~ effective action \cite{Vilkovisky84,DeWitt87,Rebhan87}
 to calculate an explicitly gauge independent $\beta$--function,
though it would still depend on the external momentum prescription.
Also a Wilson--loop approach has been used to compute a
gauge invariant
quark--antiquark potential from which an effective coupling
was defined
\cite{FujimotoY88b}. Such a definition is not directly related
to the coupling considered here.

In this paper we follow the prescription of \cite{Landsman89},
and use the \vdw~ effective action to calculate the three--gluon vertex at the
static and spatially symmetric point at momentum $\kappa$ and
at temperature
$\tau$ for a $SU(N)$ Yang--Mills gas. This approach has recently
been used in \cite{vanEijck93,vanEijckSvW93} where the
one--loop $\beta$--function was calculated and the scaling in
$\tau$ and $\kappa$ was analysed.
The choice of the static renormalization point can be partially motivated by
the fact that in the ITF it is only the zero Matsubara frequency
modes that are
soft and need resummation (see Section \ref{calc}). It also
eliminates the problem of choosing between analytic
continuations (retarded/advanced or
time/anti-time ordered) which have different soft contributions
\cite{AurenchePG92}.

Since the $\beta$--function   here is linearly related to the two--point
function by a Ward identity one might naively expect that it
would have a high temperature dependence like $\tau^2/\kappa^2$.
However, at the static point there is a
cancellation and it is found that at one
loop \cite{Landsman89,vanEijck93}
\be{betabare}
       \tau\frac{dg}{d\tau}=\frac{g^3}{8\pi^2} N\frac{21\pi^2}{16}
      \frac{\tau}{\kappa}\ .
\ee
The leading linear contribution does not come from the  hard part
of the loop integral, responsible for a $\tau^2/\kappa^2$--term,
but from soft loop momenta. Therefore,
in the spirit of the Braaten--Pisarski resummation scheme \cite{bp},
it is not consistent to stop the calculation at
the one--loop order for soft internal momenta,
but the resummed propagator and vertices must be
used to get the complete
leading contribution. The main purpose of this paper is to
perform the resummed one--loop calculation and analyse the
new result. We do not include any fermion contribution since it is
subleading at high $T$.
%
\Section{Tder}{Perturbative expansion of the
  $\beta$--function}
The RGE with the temperature and momentum $\tauk$ as
parameters was first
derived in \cite{MatsumotoNU84} using the fact that the
renormalized $n$--point
functions are formally independent of the renormalization
condition. We would
like to relate this RGE to a direct calculation of the derivative of the
three--gluon function. Let us first fix the notation and work
in the Landau
gauge in this section --- it can be shown that results in this gauge,
using the background field method, coincide with those results of the
Vilkovisky--DeWitt effective action \cite{Rebhan87}.
The inverse of the full propagator is
\bea{fullprop}
	(-i\Delta^{-1})^{ab}_{\mu\nu}&=&\delta^{ab}
(g_{\mu\nu}P^2-P_\mu P_\nu)
	-\delta^{ab}\Pi_{\mu\nu}(P)\ ,\nn
	\Pi_{\mu\nu}(P)&=&A_{\mu\nu}\Pi^T(P)+
B_{\mu\nu}\Pi^L(P)\ ,\nn
	A_{\mu\nu} &=&g_{\mu\nu}-B_{\mu\nu}-
\frac{P_\mu P_\nu}{P^2}\ ,\nn
	B_{\mu\nu}&=&\frac{V_\mu V_\nu}{V^2}\ ,\quad
	V_\mu=P^2 U_\mu-U\cdot P P_\mu\ .
\eea
The four--velocity of the heat--bath is given by $U_\mu=(1,0,0,0)$.
The Ward identities in the \vdw~ effective action are particularly
simple due to the off--shell gauge invariance, and the spatial
part of the three--gluon
vertex, for static and symmetric external momenta,
can be related to the transverse part
of the polarization tensor through
\be{threepoint}
      \Gamma^{abc}_{ijk}\tauk = g\,f^{abc}\biggl\{[g_{ij}(p-q)_k+
	g_{jk}(q-r)_i+g_{ki}(r-p)_j]
	\left(1+\frac{\Pi^T\tauk}{\kappa^2}\right)
	  + \ldots\biggr\} \ ,
\ee
where  $P_\mu=(p_0=0,\vek{p})\ ,\ p^2=\abs{\vek{p}}^2$ etc.,
$p^2=q^2=r^2=\kappa^2$, and the dots stand for
terms orthogonal to $p_i$, $q_j$ and $r_k$.
When $p_0=0$ we have $2\Pi^T=-\sum_i \Pi_{ii}$ using the
Minkowski metric.
The wavefunction and coupling constant renormalizations
($A^a_\mu\goto Z_3^{1/2} A^a_{R\mu},\ g\goto Z_3^{-1/2}g_R$)
are performed at $\tauk$ so that
\bea{RC}
	(-i\Delta^{-1}\tauk)_{ij}&=&
	\at{(\delta_{ij}p^2-p_ip_j)}{p^2=\kappa^2}\ ,\nn
	\Gamma^{abc}_{ijk}\tauk&=&
	g_R\tauk f^{abc}\at{\biggl\{[g_{ij}(p-q)_k+
	{\rm cycl.}]  + \ldots\biggr\}}{p^2=q^2=r^2=\kappa^2}\ .
\eea
We now define an effective coupling constant $g\Tk$ at another
temperature $T$ by
\be{effg}
	\Gamma^{abc}_{ijk}\Tk Z_3^{3/2}(T,\tau) \equiv
	g\Tk f^{abc}\biggl\{[g_{ij}(p-q)_k+
	{\rm cycl.}]  + \ldots\biggr\}\ ,
\ee
where $Z_3^{1/2}(T,\tau)=Z_3^{1/2}(T)/Z_3^{1/2}(\tau)$ is the
rescaling of the field which is required in
order to keep the normalization of the two--point function. In
this way $g\tauk$ measures the non--linearity of the theory. It is
now straightforward to derive
\be{btdef}
	\bt\equiv\tau\frac{dg\tauk}{d\tau}=
	-\frac{g}{2\kappa^2}\at{T\frac{d\Pi^T\Tk}{dT}}{T=\tau}\ ,
\ee
using the renormalization condition in \eq{RC}. Similarly we find
\be{bkdef}
	\bk\equiv\kappa\frac{dg\tauk}{d\kappa}=-\frac{g}{2}
	\abs{\vek{p}}\frac{d}{d\abs{\vek{p}}}
	\at{\left(\frac{\Pi^T(\tau,\abs{\vek{p}})}{\abs{\vek{p}}^2}
	\right)}
	{\abs{\vek{p}}=\kappa}\ .
\ee
For a perturbative calculation of $\bt$
($\bk$ can be treated similarly) we
need only the expression for $\Pi^T\Tk$ in the
vicinity of the renormalization
point, and we then use this in fixing the initial condition for the
RGE. If the renormalization point is chosen appropriately,
we could expect reliable results in some regime
around $(\tau,\kappa)$ from a one--loop
computation of $\Pi^T\Tk$.

We know that the leading \htl{}s are non--local and if we want
to include them through some resummation we also need non--local
counter terms.
Therefore, we write the action as
\be{action}
      \cL=-\inv{4}\tr(F^2)+\inv{2}A_\mu(-P)\pi^{\mu\nu}A_\nu(P)-
      \inv{2}A_\mu(-P)\pi^{\mu\nu}A_\nu(P)\ ,
\ee
and associate the first $\pi^{\mu\nu}$ with the ``bare''
propagator and consider
the other one as a counter term. Then, we impose on the
transverse and
longitudinal part of $\pi^{\mu\nu}$ the one--loop hard
thermal loop form
\bea{piform}
      \pi^L(p_0,p)
	&=& g^2\tauk\frac{\tau^2N}{3}\left(1-\frac{p_0^2}{p^2}\right)
	\left[1-\frac{p_0}{2p}\ln\abs{\frac{p_0+p}{p_0-p}}\right]\ ,
	\non\\[3mm]
      \pi^T(p_0,p)
	&=& g^2\tauk\frac{\tau^2N}{6}\left[\frac{p_0^2}{p^2}+
	\left(1-\frac{p_0^2}{p^2}\right)
	\frac{p_0}{2p}\ln\abs{\frac{p_0+p}{p_0-p}}\right]\ .
\eea
It is enough to introduce the momentum dependence in $\pi^{L,T}$
from \htl{}s in order to resum the leading $g^2\tau^2$ contribution.
The effective propagator, defined by
\be{geneffprop}
	(-iD^{*\,-1})^{ab}_{\mu\nu}=\delta^{ab}(g_{\mu\nu}
	P^2-P_\mu P_\nu)
	-\delta^{ab}\biggl(A_{\mu\nu}\pi^T(p_0,p)+
	B_{\mu\nu}\pi^L(p_0,p)\biggr)\ ,
\ee
 has an explicit $\tau$ dependence which leads to a
$\tau$--dependent UV divergence already at the one--loop
level, since the $iD^{*ab}_{\mu\nu}$ contains contributions from
an infinite sum of higher order diagrams. This
$\tau$--dependence disappears when all diagrams to a given
order are
included \cite{KislingerM76} and thus the problem can be
pushed to arbitrarily
high order by performing the renormalization to higher order. In
our approach we only have to assume that this had been carried
out at some $\tau$ when renormalizing $\Pi^T\tauk$. After taking the
$T$ derivative and the limit $  T\goto \tau$ everything is finite.
We also note that vertices have \htl~corrections so they
should be treated
in a similar manner by adding and subtracting the effective
vertices in \eq{action}, but we do not need them in the
approximation we
are using (see Section \ref{calc}).
\bigskip\\ ~
Let us now analyse the perturbative calculation of $\bt$
using the renormalized
Lagrangian in \eq{action}. The high
temperature expansion of $\Pi^T\Tk$ is an
expansion in $g^2T^2/\kappa^2$
and $g^2T/\kappa$. The $g^2T^2/\kappa^2$ only comes from the
\htl{}s and for each such diagram there is a corresponding
counter term $g^2\tau^2/\kappa^2$ with the opposite sign
generated by the last
term in \eq{action}. This is so because the counter term is
chosen to be
exactly the \htl~contribution.
The $\bt$--function is finally computed as the
derivative of $\Pi^T\Tk$ with
respect to $\ T$ at $T=\tau$. If a diagram
contains two or more \htl{}s the leading $g^2T^2/\kappa^2$
and $g^2\tau^2/\kappa^2$ terms factor out in such
a way that after taking the derivative and $T\goto\tau$ they
cancel. It then
follows that in the perturbative expansion of $dg\tauk/d\tau$
at most one \htl~contribute in each diagram and it is in
fact an expansion in
$g^2\tau/\kappa$ only. The cancellations
  are identical to what was found for the $\phi^4$--model in
\cite{Elmfors93} except that here we must use momentum dependent
counter terms since the \htl{}s are non--local.
Also the usual way of simply using   improved propagators to do loop
calculations, without the RGE, does indeed resum the leading
powers of
$g^2\tau^2/\kappa^2$. The difference is here that $g$ itself is
not a fixed
 zero temperature parameter but it is defined through the solution
to the RGE. Therefore, the expansion is
 really in powers of $g^2(\tau,\kappa)\tau/\kappa$ and its value
depends on the solution of the \trge. The possibility of
performing a perturbative expansion at high $\tau$ for
fixed $\kappa$
depends on whether this combination increases or decreases at
large $\tau$.
\bigskip\\ ~
Before doing any actual computation of the resummed
$\beta$--function it is
interesting to discuss what kind of new terms one can
expect and what their consequences would be.
Let us therefore write
\be{betaexp}
      \tau\frac{dg}{d\tau}=
	\beta_\tau^{(0)}+\beta_\tau^L+\beta_\tau^T=
	g^3 \frac{\tau}{\kappa}
      (c_0+c_1\frac{g\tau}{\kappa}+c_2\frac{g^2\tau}{\kappa})\ ,
\label{betaresummed}\ee
assuming a high $\tau$ expansion ($\tau\gg\kappa$).
The contribution from \htl{}s is denoted by $\beta_\tau^L$ since it is
generated by a longitudinal mass (see Section \ref{calc}).
In the expansion of $\bt$ in \eq{betabare} there is no contribution
from any \htl~and we expect that the use of resummed propagators
will supply the $c_1$ term of relative order $g\tau/\kappa$.
The inclusion of $\beta_\tau^T$ from a
transverse ``magnetic mass'' of order
$g^2\tau$, as discussed below,
would generate the $c_2$ term. We assume that
the initial condition is given at a temperature
$\tau_0\gg\kappa$ while we
still have  $g^2(\tau_0,\kappa)\tau_0
\ll\kappa$ so that we can do a consistent perturbation expansion in
$g^2\tau_0/\kappa$. As $\tau$ increases the solution to the
RGE determines
whether $g^2\tauk\tau/\kappa$ stays small enough for
the perturbative expansion
to remain valid.
With the positive sign in \eq{betabare} for the bare one--loop
$\beta$--function the coupling constant
diverges at some $\tau$ implying that the expansion
breaks down. If the sign had been negative the solution would go like
$g\tauk\sim (\tau/\kappa)^{-1/2}$, implying that
$g\tau/\kappa$ increases
and has to be resummed while $g^2\tau/\kappa$ goes to a
constant and
can be treated perturbatively if it is not too large.
In the present case the bare one--loop calculation gives a
divergent $\as$
but resummation of $g\tau/\kappa$ terms may change this.
In particular, in Eq.(\ref{betaresummed}), if $c_1$ is negative at large
$\gtk$ it dominates over the constant term and the
asymptotic form of $g\tauk$ is $(\tau/\kappa)^{-2/3}$. The
factor $\gtk$
still increases and needs to be resummed
(as done with the momentum
dependent counter terms in the \trge) but the $\gttk$  terms actually
go to zero and the exact high $\tau$ limit would be
under control. We have found (see
Section \ref{calc})
that $c_1$ is actually zero but there is
a correction to the constant $c_0$, though it does not change the sign
of $\bt$ for large $\gtk$.

It is also interesting to see what happens if a magnetic mass is
present; although such an effect is believed to be non--perturbative,
we could crudely mimic such a term perturbatively by introducing
some constant
$m_T\sim g^2\tau$ by hand as the position of the pole of the
static transverse mode. Assuming that $c_2$ is negative and
dominates we find that $g\tauk\sim (\tau/\kappa)^{-1/2}$. It may
thus be inconsistent to assume that $\gttk$  is large since it goes
to a constant, and one would have to solve the \rge~ with
the full $\tauk$ dependence. We found that $c_2$ is gauge
dependent and
positive in  the Landau gauge. Again, even if this correction
would make
$\bt$ negative it is not consistent to separate out $\gttk$ and
subleading constants since they all go to constants.

To summarize, a negative $c_1$ term would cure the problem
of a divergent perturbative expansion of $\bt$, but the
actual result shows only corrections to the $c_0$
and the sign remains positive leading to a divergent $g\tauk$ at
some finite $\tau$.
%
\Section{calc}{The resummed one--loop calculation}
To find the $\beta$--function in the scheme described in
Section \ref{intro}
we   need to compute the transverse part of the
polarization tensor
at one--loop using the effective propagators and vertices including
\htl~corrections. We shall perform the calculation in an arbitrary
covariant background field gauge for comparison with other results
and to see which terms are gauge independent, though the \vdw~
approach prescribes the Landau gauge.
The Feynman rules in the background gauge can be found in \cite{Abbott81}
and a one--loop calculation of the polarization tensor at finite $T$
was performed in \cite{ElzeHKT88}.
Let us start with $\bt$--function without resummation in a general
covariant background gauge, parametrized with $\xi$. The Landau
gauge $\xi=0$ was considered in \cite{vanEijck93,vanEijckSvW93}
and the Feynman gauge $\xi=1$ in \cite{AntikainenCPS90}.
We can extract the result for general $\xi$ from the calculation in
\cite{ElzeHKT88}. Furthermore, the leading $\tau/\kappa$ comes
from the IR
dominant part of the loop and is determined by the $n=0$
Matsubara frequency.
For diagrams that are UV--convergent we can extract the
linear $\tk$ term by
simply restricting the sum to $n=0$. Diagrams that
are not UV--convergent have
to be summed over all $n$. For the integrals we are dealing
with it turns out
that if the diagram is only logarithmically divergent it is in
fact enough to
take the $n=0$ term to get the correct leading real part.
There is an example
in \cite{ElzeHKT88} where this does not work for the imaginary part.
The expression needed for the one--loop polarization tensor
in a general
background field gauge, including the ghost
contributions, is \cite{ElzeHKT88}
(we are using a different sign convention
than \cite{ElzeHKT88})
\bea{selfenergy}
	 \Pi_{\mu\nu}(K)=\align -g^2N\,
	T\sum_n\int \frac{d^3p}{(2\pi)^3}\
	\Biggl[  g_{\mu\nu} iD^\alpha_\alpha (P)
	-\left(1-\frac{1}{\xi}\right)iD_{\mu\nu}(P)\non\\[3mm]
	\align +\,
	\frac{2P_\mu P_\nu -P_\mu Q_\nu -Q_\mu P_\nu}{P^2 Q^2}
	 -\,  \frac{2g_{\mu\nu}}{P^2}
	\non\\[3mm]
	\align+\,\frac{1}{2}  \Gamma_{\alpha\beta\mu}(P,Q,K)
	D^{\alpha\alpha'}(P) D^{\beta\beta'}(Q)
	\Gamma_{\alpha'\beta'\nu}(P,Q,K) \Biggr]\ ,
\eea
where $P+Q+K=0$ and the bare three--point vertex is
\begin{equation}
	\Gamma_{\alpha\beta\mu}(P,Q,K)=g_{\alpha\beta}
	\left(P-Q\right)_\mu
	+g_{\beta\mu}\left(Q-K+\frac{1}{\xi}P\right)_\alpha
	 +g_{\mu\alpha}\left(K-P-\frac{1}{\xi}Q\right)_\beta.
\end{equation}
Calculating the transverse function $\Pi^T(p_0=0,p^2=\kappa^2)$
we find using the bare propagators in Eq.(\ref{selfenergy}) the
following result for $\beta_\tau^{(0)}$ of Eq.(\ref{betaresummed}):
\be{bbgeng}
	\beta_\tau^{(0)}=\frac{g^3}{8\pi^2}\frac{N\pi^2}{16}(21+6\xi+
	\xi^2)
	\frac{\tau}{\kappa}\ .
\label{beta0}\ee
For $\xi=0$ and $\xi=1$ this
coincides with  \cite{vanEijck93} and  \cite{AntikainenCPS90},
respectively, and
confirms the conjecture in  \cite{vanEijckSvW93} that the difference
between their result and that of
 \cite{AntikainenCPS90} is due to the gauge choice.

The general one--loop calculation with effective propagators
and vertices is
difficult but in our case there are some simplifications.
First we consider the external energy to be zero, and in the
ITF only the $n=0$
internal modes need resummation since all other modes are hard.
The effective
propagators are thus only needed for zero energy and then they
take the simple form
\be{effprop}
	D^{*ab}_{\mu\nu}(0,\vek{p})=-i\delta^{ab}
	\left[-\inv{p^2+m_T^2}(-\delta_{ij}+\frac{p_ip_j}{p^2})-
	\inv{p^2+m_L^2}\delta_{\mu 0}\delta_{\nu 0}+
	\xi\frac{p_ip_j}{p^4}\right]\ ,
\ee
where $m_T$ and $m_L$ are transverse (magnetic) and
longitudinal (electric)
masses respectively. The longitudinal electric mass
$m_L^2=\frac{1}{3} g^2N\tau^2$ comes from the one--loop
hard thermal
loops, but the transverse magnetic mass $m_T$ is zero perturbatively.
Here we try to estimate its effects by hand by inclusion
of the term $m_T\sim O(g^2\tau)$
in the propagator as a crude approximation
to the true (non--perturbative)
situation.
Only the pure gauge boson diagrams are effected by
the resummation. The
correction to e.g. the tadpole diagram is
\be{taddiff}
	\frac{T}{2}\sum_n\int\dk{3}
	\biggl(\gamma^{*abcd}_{\mu\nu\alpha\beta}
	D^{*cd}_{\alpha\beta}(K)-
	\gamma^{abcd}_{\mu\nu\alpha\beta}
	D^{cd}_{\alpha\beta}(K)\biggr)\ ,
\ee
where the star $*$ denotes effective vertices and propagators.
Each of the two
terms in \eq{taddiff} is quadratically divergent and
receives contribution from
all Matsubara frequencies at high $T$. The difference however is only
logarithmically divergent and to get the leading $\tk$ term
only the $n=0$ mode is
needed. It then follows that
$\gamma^{*abcd}_{\mu\nu\alpha\beta}$ is only
needed for zero external energy and then it reduces to the
bare vertex.
Similar simplifications can be done for the bubble diagram.
The degree of
divergence is reduced by two when subtracting the
unresummed result and that is
enough for using the $n=0$ approximation. We write the
additional contribution
to $\Pi^T$ from non--zero $m_L$ and $m_T$ like
\be{delPi}
	\delta\Pi^T(z_L,z_T)=\delta_L\Pi^T(z_L)
	+\delta_T^{(\xi=1)}\Pi^T(z_T)
	+(1-\xi)\delta_T^{(\xi\neq1)}\Pi^T(z_T)\ ,
\ee
where $z_L=m_L/\kappa$ and $z_T=m_T/\kappa$. The
explicit expressions turn out to be
\bea{delPiexpl}
	\delta_L\Pi^T&=& \frac{g^2N}{4\pi^2} T \kappa
	\left\{
	-\frac{\pi}{2}z_L + \frac{\pi^2}{2}z_L^2
	-\frac{\pi}{4}[1+4z_L^2]\arctan(2z_L)\right\}
	\ ,\nn
	 \delta_T^{(\xi=1)} \Pi^T&=&
	\frac{g^2N}{4\pi^2} T \kappa
	\left\{-\frac{\pi}{8}\frac{5+4z_T^2}{z_T}+\frac{3\pi^2}{4}z_T^2\right.
	\nonumber\\
	& & \left.~~~~~~~~~~~~
	-\frac{\pi}{16}\frac{ (4z_T^2+1)(8z_T^4-12z_T^2+1)}{z_T^4}
	\arctan(2z_T) \right.
	\nonumber\\
	& & \left.~~~~~~~~~~~~
	 + \frac{\pi}{8}\frac{4z_T^6+3z_T^4-4z_T^2+1}{z_T^4}
	\arctan(z_T)\right\}
	\ ,\nonumber\\
	\delta_T^{(\xi\neq1)} \Pi^T&=&
	\frac{g^2N}{4\pi^2} T \kappa
	\left\{ -\frac{\pi}{4}\frac{2z_T^2-1}{z_T}-\frac{\pi^2}{4}z_T^2
	-\frac{\pi}{4}\frac{1+z_T^2-2z_T^4}{z_T^2}
	\arctan (z_T)\right\}\ .
\eea
In this the following integrals have been used:
\bea{Piint}
	  \int_0^\infty \frac{dx}{x}\ln\left(\frac{x+1}{x-1}\right)^2
	&=&\pi^2\ , \nonumber\\
	  \int_0^\infty \frac{x\,dx}{x^2+z^2}
	\ln\left(\frac{x+1}{x-1}\right)^2
	&=&\pi^2-2\pi\arctan (z)\ , \nonumber\\
	  \int_0^\infty \frac{dx}{x}\ln
	\left[\frac{(x+1)^2+z^2}{(x-1)^2+z^2}\right]
	&=&\pi^2-2\pi\arctan (z)\ , \nonumber\\
	  \int_0^\infty \frac{x\,dx}{x^2+z^2}\ln
	\left[\frac{(x+1)^2+z^2}{(x-1)^2+z^2}\right]
	&=&\pi^2-2\pi\arctan (2z)\ .
\eea
The limit of $\delta\Pi^T$ for small $z$ is given by
\bea{smallz}
	 \delta_L\Pi^T&\simeq&
	\frac{g^2N}{4\pi^2} T \kappa \left[
	-\frac{\pi}{4}z_L+\frac{\pi^2}{8}z_L^2-\frac{\pi}{3}z_L^3+\ldots
	\right]\ ,\nonumber\\
	 \delta_T^{(\xi=1)} \Pi^T&\simeq&
	\frac{g^2N}{4\pi^2} T \kappa \left[
	\frac{10\pi}{3}z_T+\frac{3\pi^2}{4}z_T^2-\frac{91\pi}{15}z_T^3+\ldots
	\right]\ ,\nonumber\\
	 \delta_T^{(\xi\neq 1)} \Pi^T&\simeq&
	\frac{g^2N}{4\pi^2} T \kappa \left[
	-\frac{2\pi}{3}z_T-\frac{\pi^2}{4}z_T^2+\frac{8\pi}{15}z_T^3+\ldots
	\right]\ ,
\eea
while for large $z$ we find
\bea{largez}
	 \delta_L\Pi^T&\simeq&
	\frac{g^2N}{4\pi^2} T \kappa \left[
	-\frac{\pi^2}{8}+\frac{\pi}{12 z_L}-\frac{\pi}{240 z_L^3}+\ldots
	\right]\ ,\nonumber\\
	 \delta_T^{(\xi=1)} \Pi^T&\simeq&
	\frac{g^2N}{4\pi^2} T \kappa \left[
	\frac{23\pi^2}{16}-\frac{13\pi}{6z_T}+\frac{47\pi}{120z_T^3}+\ldots
	\right]\ ,\nonumber\\
	 \delta_T^{(\xi\neq 1)} \Pi^T&\simeq&
	\frac{g^2N}{4\pi^2} T \kappa \left[-\pi z_T -
	\frac{\pi^2}{8}+\frac{2\pi}{3z_T}-\frac{\pi^2}{8z_T^2}+\ldots
	\right]\ .
\eea
It is worth noting that $\delta_L\Pi^T$ is independent of the
gauge parameter $\xi$ and that it contains terms that are
potentially dominant
for large $z_L$. However, it turns out that the leading terms
 cancel between
the tadpole and the bubble diagram, and that $\delta_L\Pi^T$
only contributes
to $c_0$ (and not to $c_1$) in Eq.(\ref{betaresummed}).
 When added to the bare one--loop result
$\beta_\tau^{(0)}$
of Eq.(\ref{beta0}) we find for $\xi=0$
\be{blandau}
	\beta_\tau^{(0)}+\beta_\tau^L
	\simeq\frac{g^3}{8\pi^2} N\frac{\tau}{\kappa}
	\frac{23\pi^2}{16}\ .
\ee
The resummation has not changed the sign but the
quantitative result to this order, showing that it was necessary
to include these effects for a consistent calculation.
The results of \cite{Landsman89,vanEijck93,vanEijckSvW93}
are in this sense incomplete,
but the general conclusions are correct
since the sign remains unchanged.
They could have been drastically changed if, for
instance, the linear $m_L$ had come non--vanishing and
negative (see Section \ref{Tder}).

When including $m_T$ the large $z$ limit gives
\be{bresum}
	\beta_\tau\simeq\frac{g^3}{8\pi^2} N\frac{\tau}{\kappa}\left(
	\frac{(\xi+3)^2+12}{16}\pi^2+\frac{\pi^2}{8}+
	\left[(1-\xi)(\pi \frac{m_T}{\kappa}+\frac{\pi^2}{8})
	-\frac{23}{16}\pi^2\right]\right)\ ,
\ee
where everything inside  the [\ ]--parentheses comes from
the inclusion of
$m_T$. In the Landau gauge $\bt$ is positive even when
including $m_T$.
Even though it is possible to choose a gauge with a
large enough $\xi$ in
order to
stabilize the running of $\as$ it seems rather artificial since
we only can
argue in favour of the $\xi=0$ gauge from the \vdw~ approach.

We have numerically solved the RGE in the high $\tau$
limit at a fixed momentum scale $\kappa$
(i.e. neglecting the vacuum contribution and expanding in
$\kappa/\tau$) but using the
exact dependence on $m_L$ and $m_T$. The result is
presented in \figs{fig-xi02}{fig-lt}.
\bfig
	\epsfxsize=14cm
	\epsfbox{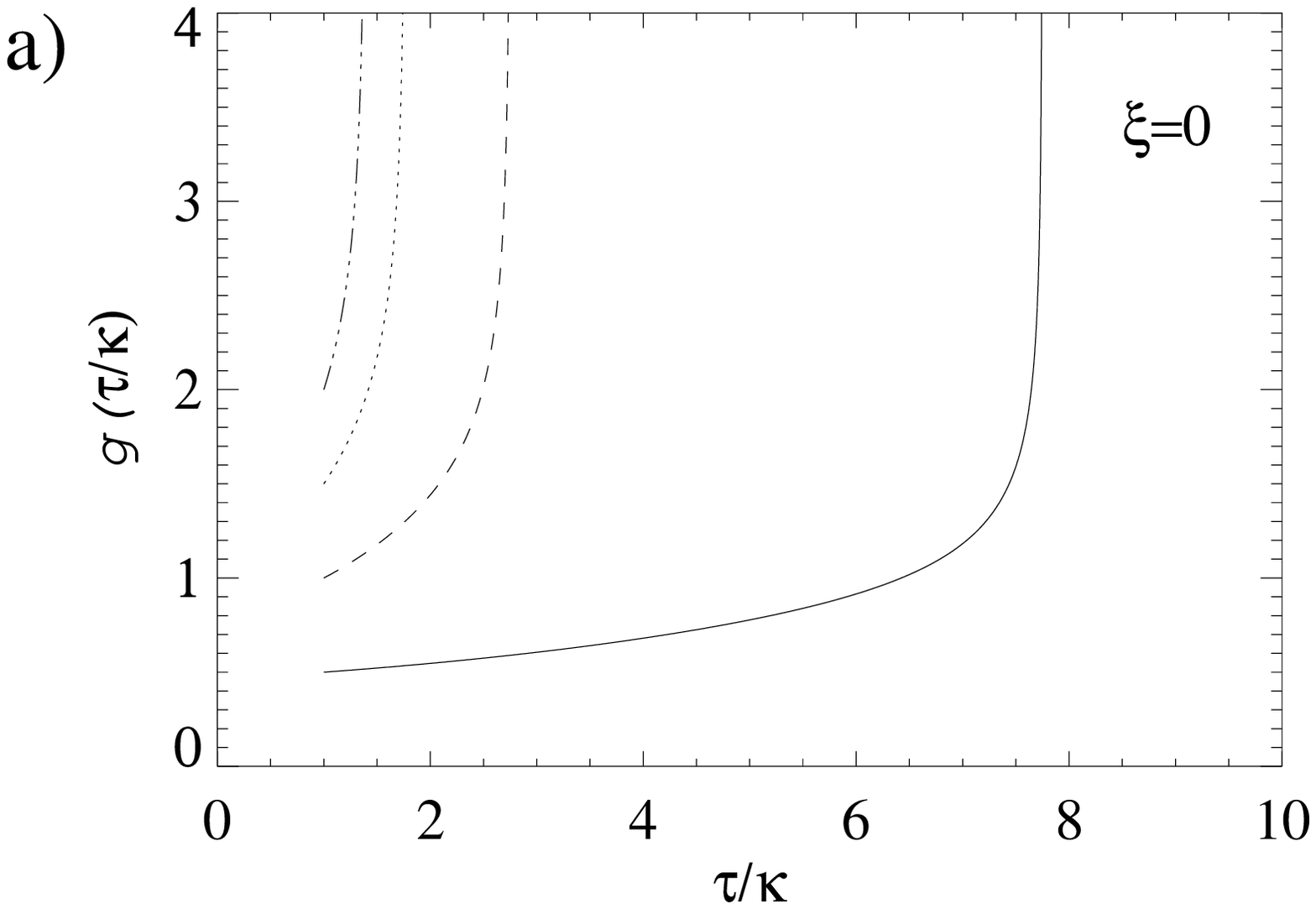}
	\epsfxsize=14cm
	\epsfbox{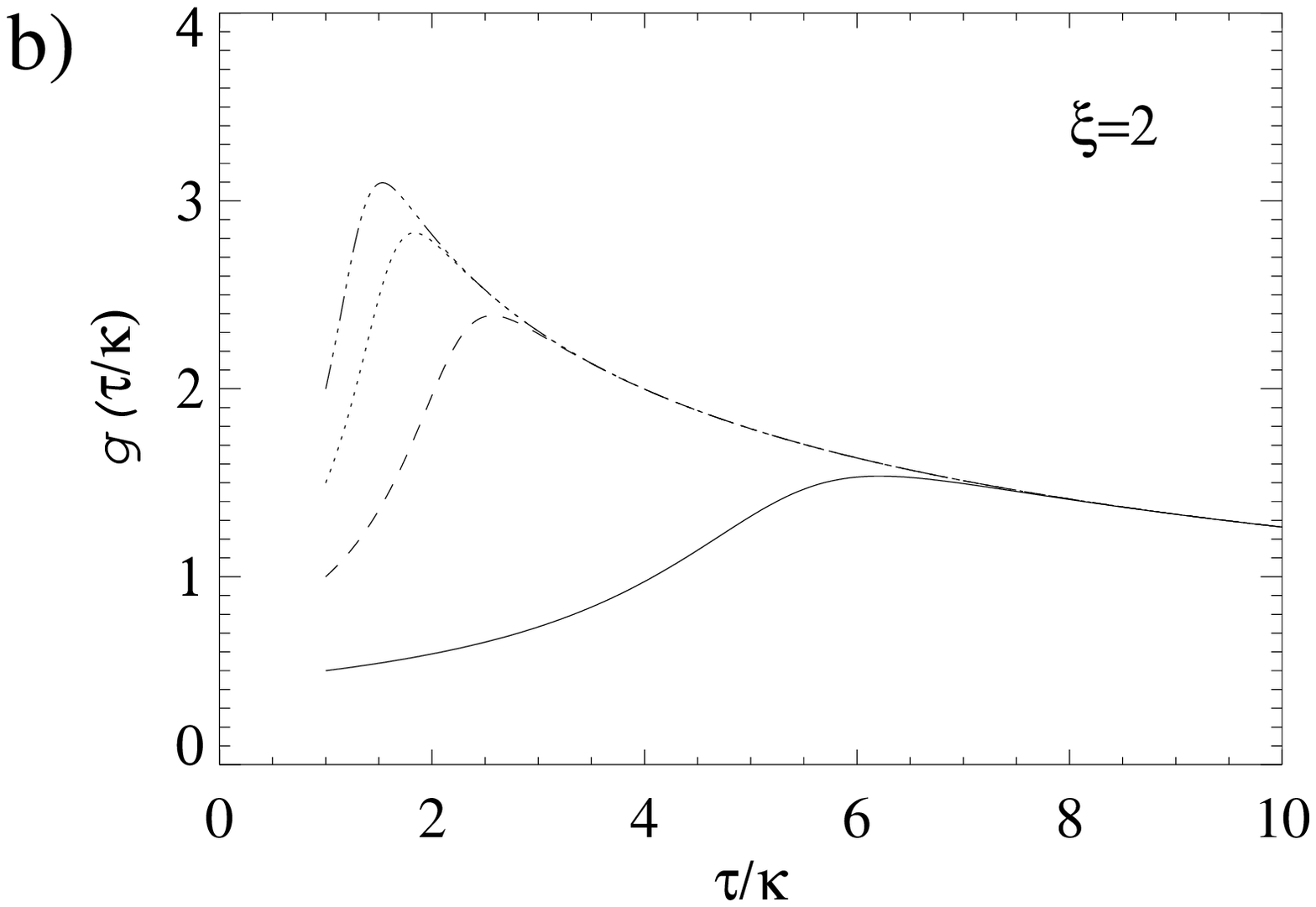}
\baselineskip 13pt
\figcap{The running coupling constant in a $SU(2)$ Yang--Mills
theory for the initial conditions $g(1)=0.5,\ 1.0,\ 1.5$ and $2.0$.
Only the thermal contribution is included. The hard thermal loops
are resummed and a magnetic mass $m_T=c g^2 \tau$ with $c=0.24$,
is included. The value of $c$ is taken from the numerical
simulations in \cite{BilloireLS81}. In the Landau gauge,
$\xi=0$ (Fig.(1a)),
the coupling diverges at a finite temperature and
perturbation
theory breaks down. In the $\xi=2$ gauge (Fig.(1b))
the contribution from non--zero $m_T$ prevents the coupling
from diverging.}
\label{fig-xi02}
\efig
In the $\xi=0$ gauge (Fig.(1a)) the coupling constant diverges at
a finite temperature, just like without resummation. If we
choose $\xi>1$, e.g. $\xi=2$ as  in Fig.(1b), the contribution
from $m_T$ changes the sign of $\bt$ for large $\tau/\kappa$.
\bfig[t]
	\epsfxsize=15cm
	\epsfbox{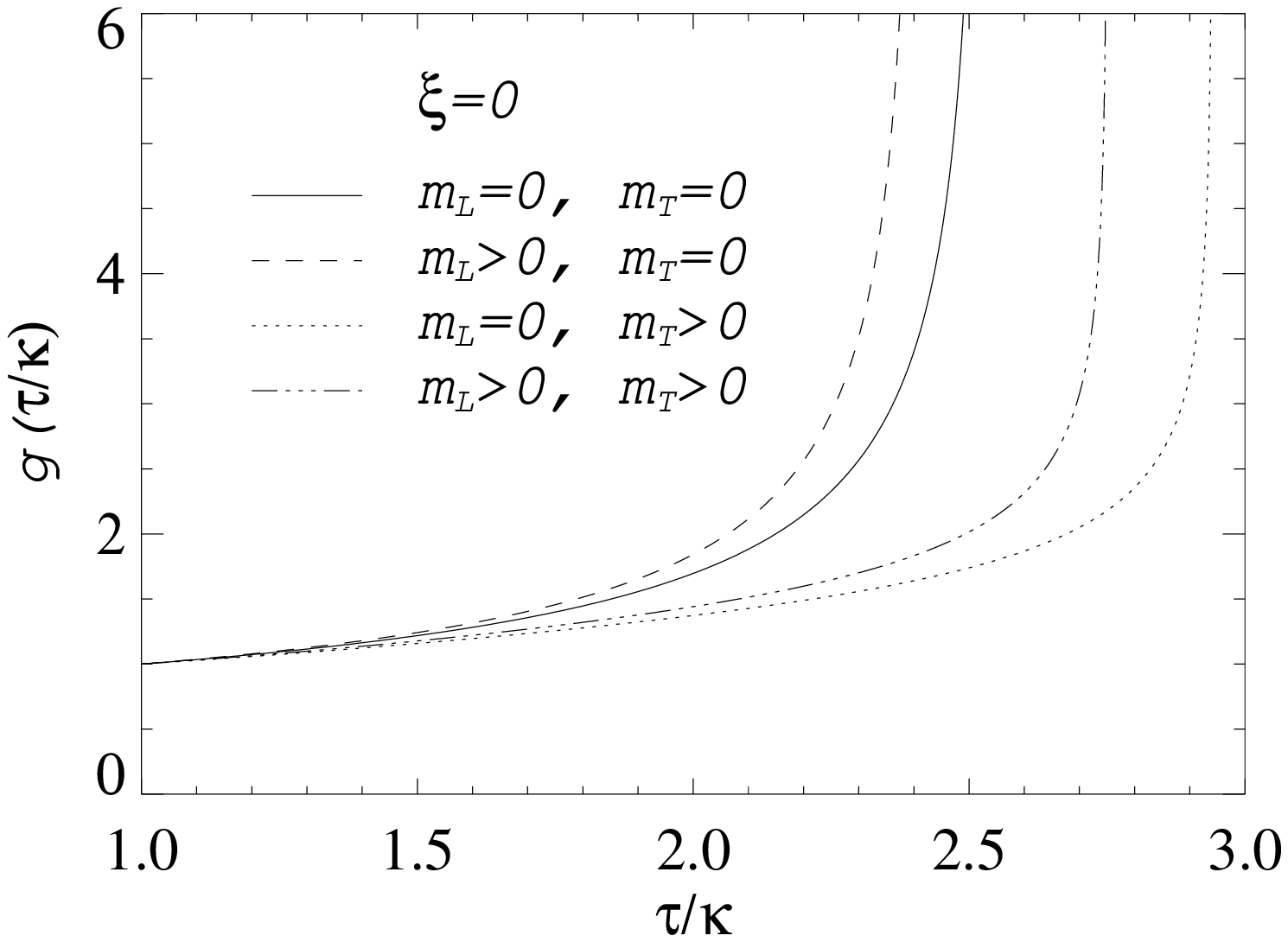}
\baselineskip 13pt
\figcap{The effects of including or excluding $m_L$ and $m_T$
in the $\bt$-function. The parameters are the same as in
\fig{fig-xi02} with $\xi=0$ and $g(1)=1$.}
\label{fig-lt}
\nopagebreak
\efig
To see the effect of resummation in the $\xi=0$ gauge we have
computed $g(\tk)$  with and without the contribution from $m_L$
and $m_T$ (see \fig{fig-lt}). We find that the qualitative behaviour
is not drastically changed by the resummation. The inclusion of $m_L$
has a tendency to increase the growth of $g(\tk)$ while $m_T$ pushes
the divergence to a higher temperature.
%
\Section{disc}{Discussion}
%
The problems associated with a consistent
calculation to this order of the $\beta$--function
concerning the ``wrong sign'' and the gauge dependence are
reminiscent of the early one--loop bare calculations of the
gluon damping constant at rest. Such calculations also gave
the ``wrong sign'', in that the modes were anti--damped,
and the results were also gauge parameter dependent. The use
of the Vilkovisky--DeWitt effective action to address the problem
of gauge dependence in this case did not resolve the problem
completely, as the damping constant calculated in this formalism
still had the wrong sign. Indeed, there were also arguments for
choosing the background field Feynman gauge $\xi=1$ as the
``preferred'' gauge based on the gauge invariant propagator
of Cornwall \cite{Nadkarni89}, but this too gave the
wrong sign for the bare one--loop
damping constant and was quantitatively different than the
Vilkovisky--DeWitt gauge $\xi=0$.
 The resolution to these
problems was later supplied by the
Braaten--Pisarski resummation scheme \cite{bp},
where a gauge invariant
and positive result is obtained to first order.
Lessons from this could be
drawn for this calculation of the $\beta$--function. The results
presented here indicate that higher--loop
effects can change the result quantitatively, but the particular
corrections considered here were not enough
to resolve the problems of the wrong sign and of gauge
dependence. This may mean that if the renormalization
group equations in this form are to provide a useful
tool a further resummation
below the soft $O(gT)$ scale is needed to do a
consistent calculation for the $\beta$-function; as argued in
 \cite{vanEijckSvW93}, the fact that the combination
$\kappa/\tau$ appears means that a large temperature
expansion is in a sense the same as probing the infrared
behaviour. The need for such a further resummation in this
context can also be seen when the simultaneous
running of the coupling
constant with temperature $\tau$ and
momentum scale $\kappa$ is investigated;
from Eqs.(\ref{btdef},\ref{bkdef}), we see that with the
particular resummation investigated here
the integrability condition
\be{intblty}
	\tau\frac{d}{d\tau}\bk=\kappa\frac{d}{d\kappa}\bt\
\ee
is not automatically satisfied. This particular problem could be
cured in a somewhat {\sl ad hoc} manner by having the
$T$--derivative in Eq.(\ref{btdef}) act not only on the explicit
$T$ dependence arising from the Matsubara frequency sum
but also on the implicit
$T$ dependence of the masses $m_L$ and $m_T$.
Doing so changes the results presented here only
slightly quantitatively, however. One might thus expect
that an improved resummation scheme, as well as addressing
the problems of the sign of the $\beta$--function and of
gauge dependence,
would also yield an integrable set of equations for
$g(\tau,\kappa)$.
The need for such a resummation beyond that of
Braaten--Pisarski
has also been recognized in the calculations of the damping
rates of moving particles and of the production rates of
soft photons \cite{Baier94}. Whether such a scheme can be
developed and
can be used to give tractable results remains to be seen.
\bigskip\\ ~
{\bf Acknowledgments}
\medskip\\~
We are grateful for the hospitality at ENSLAPP, Annecy,
where most of this work was carried out, and for
stimulating discussions with the staff and visitors,
especially P.~Aurenche. P.~E.~wishes to thank
The Swedish Institute for financial support. R.~K.~thanks NSERC of
Canada and CNRS of France for financial support.
%

\end{document}